\documentclass{webofc}
\usepackage[varg]{txfonts}   
\usepackage{subfigure} 
\newcommand{\eps}{\epsilon}

\renewcommand{\th}{\theta}   

\newcommand{\lsim}{\, \, \raisebox{-0.8ex}{$\stackrel{\textstyle <}{\sim}$ }}

\newcommand{\beq}{\begin{equation}}
\newcommand{\eeq}{\end{equation}}
\newcommand{\ba}{\begin{array}}
\newcommand{\ea}{\end{array}}
\newcommand{\bea}{\begin{eqnarray}}
\newcommand{\eea}{\end{eqnarray}}
\newcommand{\bi}{\begin{itemize}}  
\newcommand{\ei}{\end{itemize}}
\newcommand{\ben}{\begin{enumerate}} 
\newcommand{\een}{\end{enumerate}}
\newcommand{\bc}{\begin{center}}
\newcommand{\ec}{\end{center}}

\newcommand{\txt}{\textstyle}

\newcommand{\third}{{\txt \frac{1}{3}}}

\newcommand{\Tr}{{\rm Tr}}
\newcommand{\Lagr}{\mathcal{L}}

\usepackage[normalem]{ulem}  

\begin{document}
\title{Vortex structure in superfluid color-flavor locked quark matter}
%
%

\author{\firstname{Mark G.} \lastname{Alford}\inst{1}\fnsep\thanks{\email{alford@physics.wustl.edu}} \and
        \firstname{S.~Kumar} \lastname{Mallavarapu}\inst{1}\fnsep\thanks{\email{kumar.s@wustl.edu}} \and
        \firstname{Tanmay} \lastname{Vachaspati}\inst{2}\fnsep\thanks{\email{tvachasp@asu.edu}} \and
        \firstname{Andreas} \lastname{Windisch}\inst{1}\fnsep\thanks{\email{windisch@physics.wustl.edu}}
}

\institute{Physics Department, Washington University, St. Louis, MO 63130, USA 
\and
           Physics Department, Arizona State University, Tempe, AZ 85287, USA 
          }

\abstract{
The core region of a neutron star may feature quark matter in the color-flavor-locked (CFL) phase. 
The CFL condensate breaks the baryon number symmetry, such that the phenomenon of superfluidity arises. 
If the core of the star is rotating, vortices will form in the superfluid, carrying the quanta of angular momentum. 
In a previous study we have solved the question of stability of these vortices, where we found numerical proof 
of a conjectured instability, according to which superfluid vortices will decay into an arrangement of 
so-called semi-superfluid fluxtubes. Here we report first results of an extension of our 
framework that allows us to study multi-vortex dynamics. 
This will in turn enable us to investigate the structure of semi-superfluid string lattices, 
which could be relevant to study pinning phenomena at the boundary of the core.
}
\maketitle
%
\section{Introduction}
\label{intro}
\begin{figure}
\bc
\includegraphics[width=0.5\hsize]{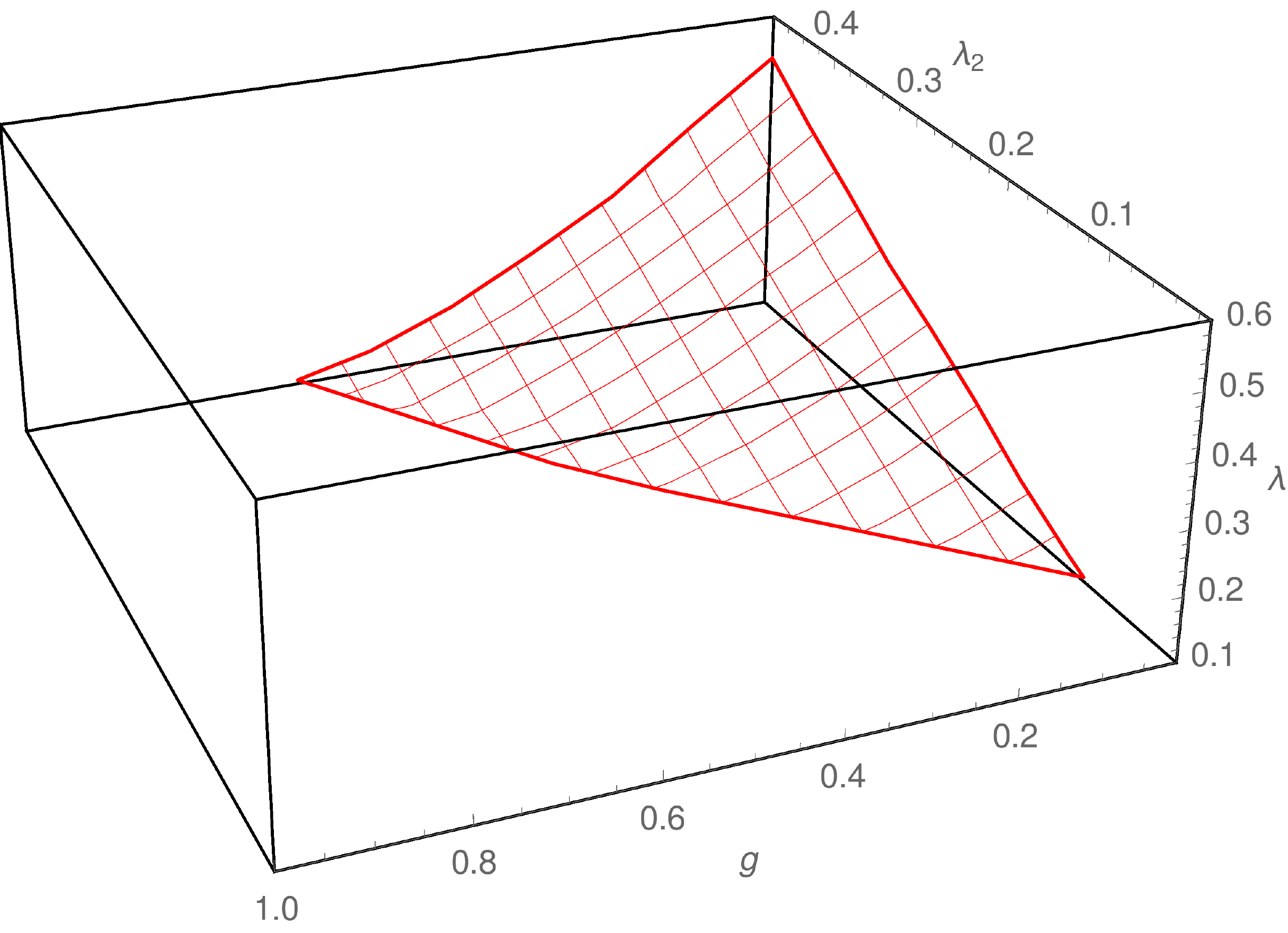}
\caption{The parameter space of the couplings $g$, $\lambda$ and $\lambda_2$. Points behind the surface form the region of meta-stability, all other points constitute the region of instability.}       
\label{fig:parameter_space_scan_plot} 
\ec 
\end{figure}

\begin{figure*}
\centering
\subfigure[][]{
 \label{fig:parameter_space_contour_plot_a}
\includegraphics[width=0.40\hsize]{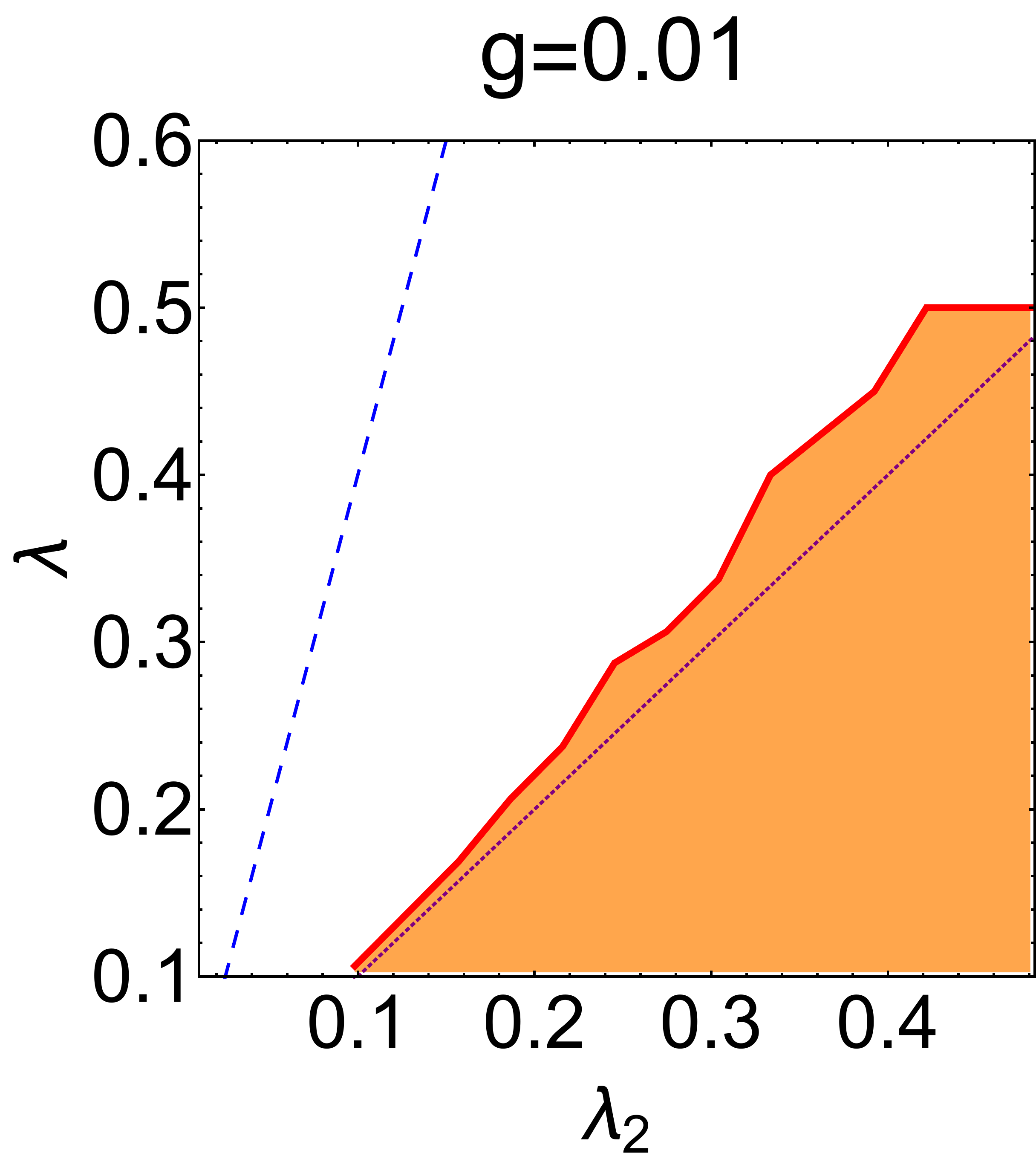}
}\hspace{8pt}
\subfigure[][]{
 \label{fig:parameter_space_contour_plot_b}
\includegraphics[width=0.40\hsize]{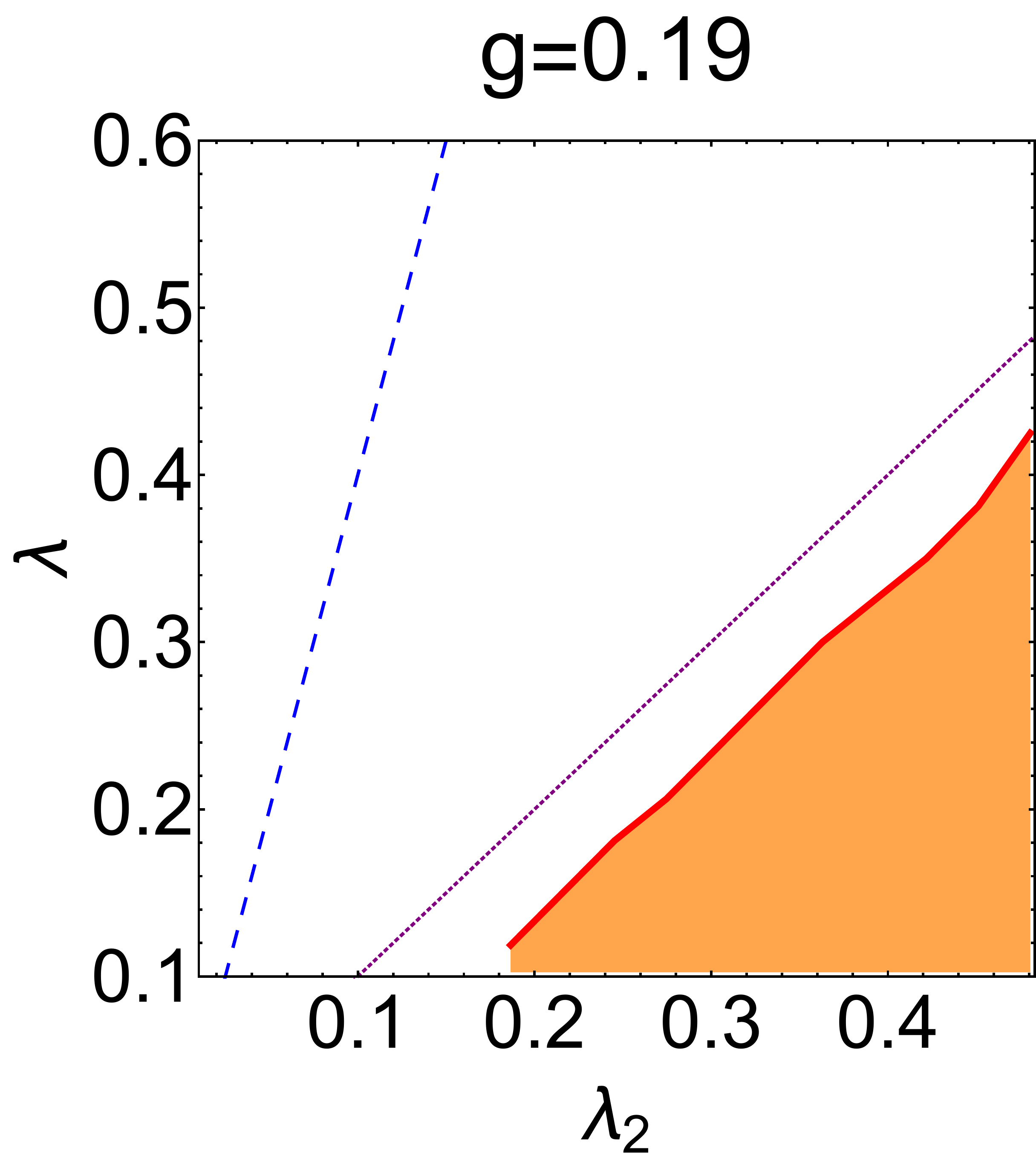}
}\\
\subfigure[][]{
 \label{fig:parameter_space_contour_plot_c}
\includegraphics[width=0.40\hsize]{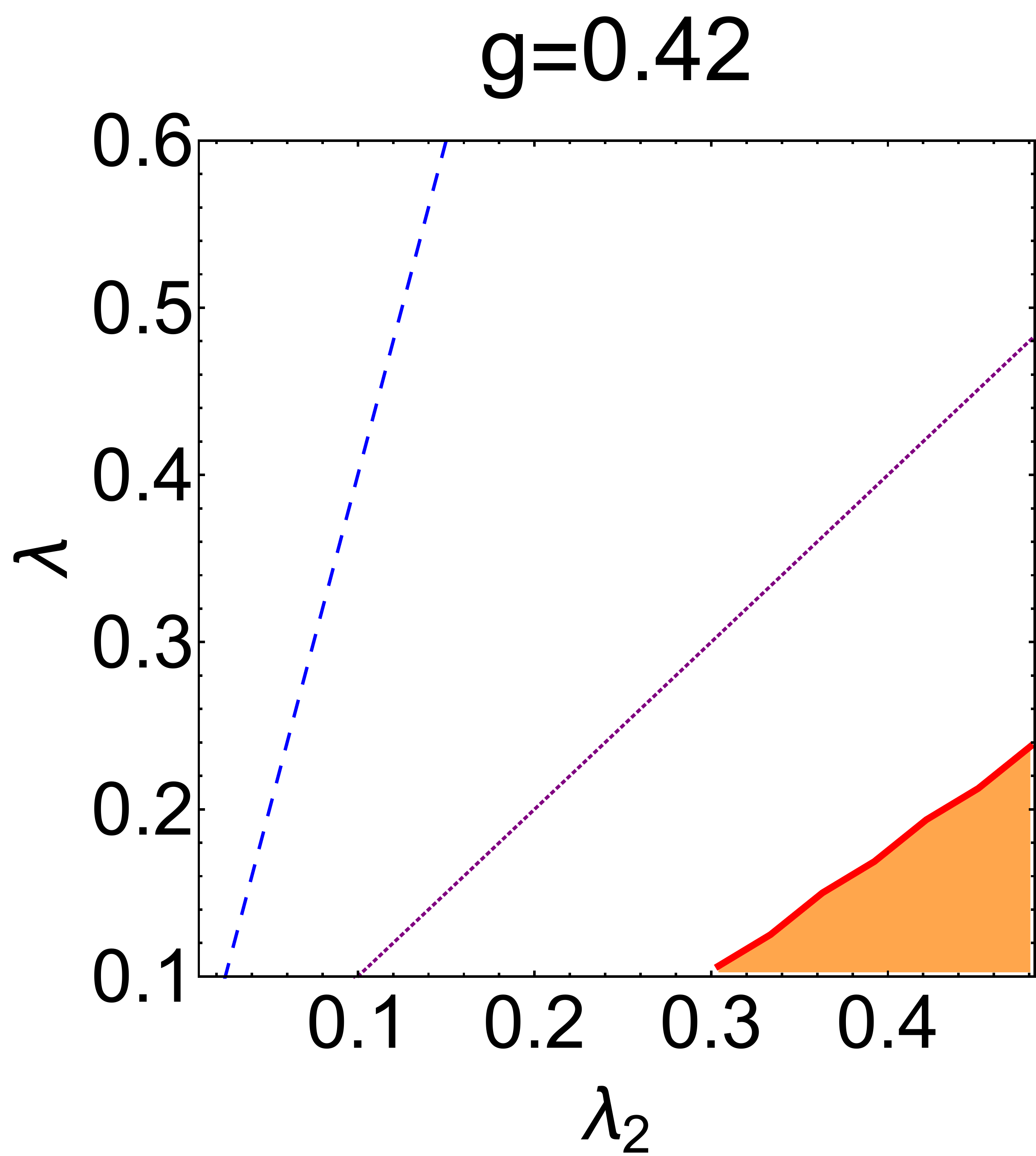}
}
\hspace{8pt}
\subfigure[][]{
 \label{fig:parameter_space_contour_plot_d}
\includegraphics[width=0.40\hsize]{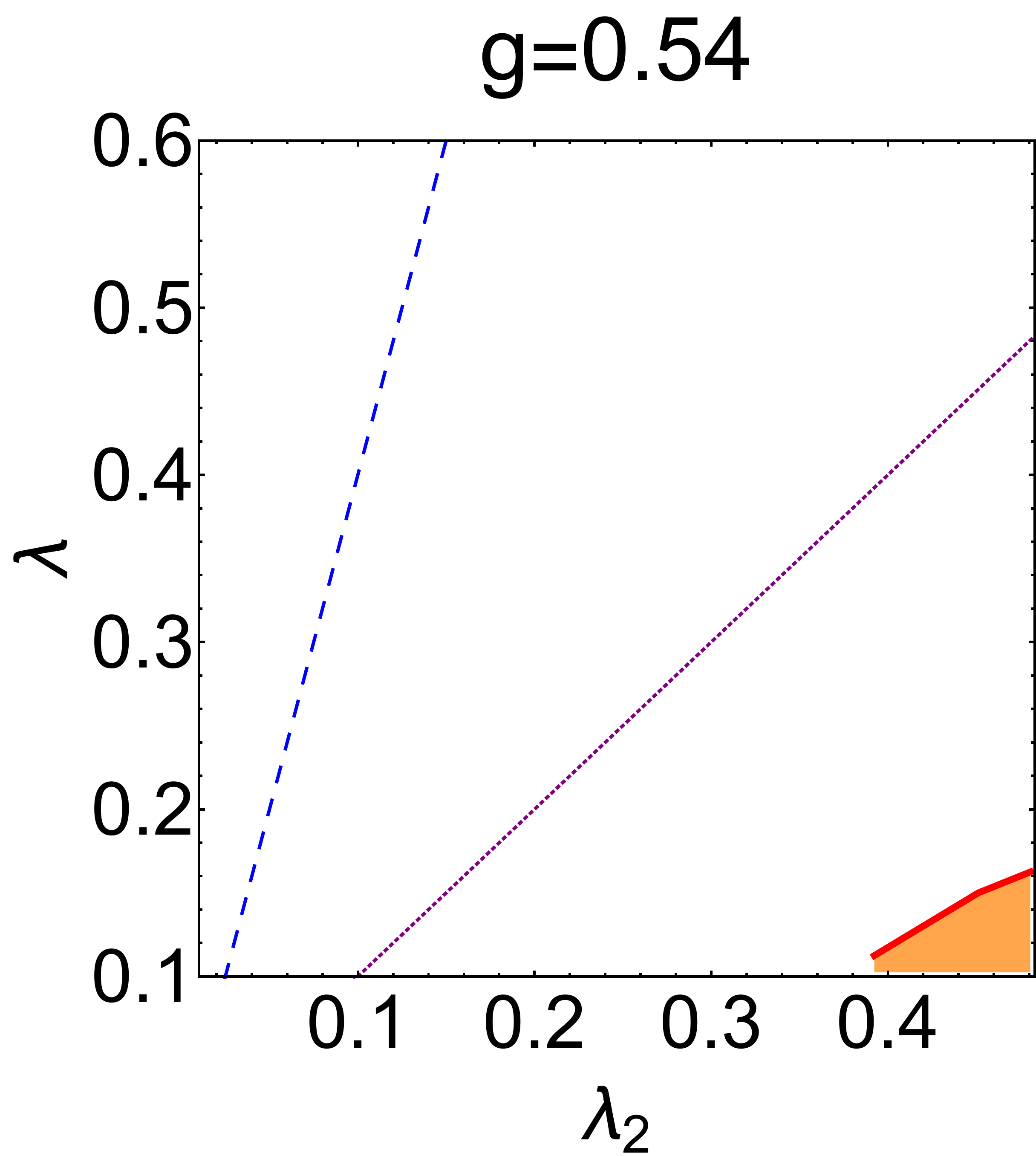}
}
\caption[]{Slices through the parameter space of the couplings for different values of $g$. The shaded area is the region of meta-stability, its complement constitutes regions of instability. The dashed line is the projection of the plane of weak-coupling results, see text. The solid line represents $\lambda=\lambda_2$, which, in the case of vanishing gauge coupling, should yield the boundary of stability/meta-stability. Note that the metastability region can be characterized by $\lambda_1\lsim -0.16g$. For fixed values of $g$ and $\lambda_1$, a change in $\lambda_2$ has almost no effect on the stability boundary.}
\label{fig:parameter_space_contour_plot}
\end{figure*}

In its densest form, matter appears in the color-flavor locked (CFL) phase \cite{Alford:1997zt}. 
The CFL condensate breaks the baryon number symmetry, which renders this 
phase a superfluid. If this form of matter is present in the core region of a 
rotating neutron star, vortices will carry the angular momentum of the spinning core.
In a recent study \cite{Alford:2016dco}, we addressed the question 
of stability of these superfluid vortices, using a Ginzburg-Landau
effective theory. There, from a topological point of view, stable  vortex solutions are 
expected, since the first homotopy group of the vacuum manifold is non-trivial \cite{Balachandran:2005ev}.
However, the global vortex solution does \textit{not} possess the lowest energy,
and it has been conjectured that the vortex undergoes a decay into a configuration 
of a triplet of well-separated semi-superfluid strings \cite{Nakano:2007dq}.
In our study \cite{Alford:2016dco}, we not only observed and numerically confirmed this decay, 
but we also mapped out the stability-/metastability boundary in the parameter space of the couplings.
We furthermore identified an analytically constructed mode that proved to be sufficient to
trigger the decay of a global vortex. Let us briefly review our findings here. 
Assuming $m_u=m_d=m_s=0$, the Ginzburg-Landau Lagrangian of the effective theory reads
\beq \label{Lagrangian}
\Lagr = \Tr \left[-\frac{1}{4} F_{ij} F^{ij} + D_{i} \Phi ^{\dagger} D^{i} \Phi   + m^2 \Phi ^{\dagger} \Phi  - \lambda_{2} (\Phi ^{\dagger} \Phi)^2 \right] - \lambda_{1}(\Tr[\Phi^{\dagger} \Phi])^2 + \dfrac{3m^{4}}{4\lambda}\ ,
\eeq
where $D_{i} = \partial_{i} - i g A_{i}$ is the covariant derivative, $F_{ij} = \partial_{i} A_{j} - \partial_{j} A_{i} - ig  \left[ A_{i},A_{j} \right] $ is the gauge field-strength tensor, and the  $A_{i}$ represent the gauge fields (gluons).
The coupling $\lambda$ is a linear combination of the original self-couplings of the condensate,
\beq \label{eq:lambda_def}
\lambda \equiv 3 \lambda_{1} + \lambda_{2}.
\eeq
The matter field $\Phi$ represents the CFL condensate, and has a $3_c\times3_f$ complex matrix structure. A general entry of the $\Phi$ matrix is thus characterized by a color index $\alpha$ and a flavor index $a$, $\phi_{\alpha a}$. 
In the broken phase, the vacuum expectation value (vev) is given by
\beq \label{eq:vev}
A_i = 0  \ , \quad  \Phi = \bar \phi \textbf{1}_{3 \times 3} \ ,
\quad \bar\phi = \sqrt{\frac{m^2}{2 \lambda}} \ .
\eeq
A (global) \textit{superfluid vortex} can then be written as
\beq \label{eq:superfluid_solution}
A_i = 0  \ ,\quad
 \Phi_{\mathrm{sf}} = \bar\phi\, \beta(r) e^{i \theta} \,\textbf{1}_{3 \times 3} \ , 
\eeq
where $\beta(r)$ is the radial profile obtained from solving
\beq \label{eq:radial_profile}
\beta^{''} + \frac{\beta^{'}}{r} - \frac{\beta}{r^{2}} - m^{2} \beta (\beta^{2} - 1) = 0 \ ,
\eeq
with the boundary conditions $\beta \rightarrow 0 \ \mathrm{as} \ r \rightarrow 0 \ $, and $\beta \rightarrow 1 \ \mathrm{as} \ r \rightarrow \infty \ $.
A (red) \textit{semi-superfluid flux tube}, on the other hand, can be written as 
\beq \label{SemisuperfluidSoln}
\Phi_{\rm ssft}(r, \theta) = \bar\phi\,
  \left(\ba{ccc}f(r){\rm e}^{i \theta} & 0 & 0 \\
  0 & g(r) & 0 \\
  0 & 0 & g(r) \ea \right),
\eeq
\beq
A^{\rm ssft}_{\theta}(r) = - \dfrac{1}{gr} ( 1-h(r)) \left(\ba{ccc}-\frac{2}{3} & 0 & 0 \\
  0 & \frac{1}{3} & 0 \\
  0 & 0 & \frac{1}{3} \ea \right),
\eeq
\beq
A^{\rm ssft}_{r} = 0 \ .
\eeq
and the solutions for the green and blue flux tubes follow from swapping the diagonal elements of the matrix.
The profile functions $f(r)$, $g(r)$ and  $h(r)$ obey a set of coupled differential equations, see equations (10)-(14) in \cite{Alford:2016dco}. As discussed in our paper, far from the core of the vortex, the energy density of one semi-superfluid flux tube is one ninth of the energy density of a global vortex,
\beq
\ba{rcl}
\eps_{\rm sf} &=& 3 \bar\phi^2/r^2,  \\[1ex]
\eps_{\rm ssft} &=& \third \bar\phi^2/r^2,  
\ea
\label{eq:edens}
\eeq
which is the cause of the instability.
In the case of vanishing gauge coupling, we identified an unstable mode analytically,
\beq
\ba{rcl}
\delta \Phi^{(8)} &=&  \eps\, \hat n\!\cdot\!\nabla \psi(r,\th)\, T_8 \ . \\[1ex]
\ea
\label{eq:unstable-mode}
\eeq
This mode establishes a distortion of the red and green
components of the vortex by a small amount $\eps$
in direction of the unit vector $\hat n$, while shifting 
the blue component by an amount of  $2\eps$ in the opposite direction.
This perturbation of the global vortex can be plugged into the Hamiltonian density, and the change in energy density evaluates to
\beq\label{eq:prediction}
\delta E_{8} =  \epsilon^{2} (\lambda_{2} -\lambda) \dfrac{\pi m^4 }{\lambda^{2}} \int_{0}^{\infty} \! r \mathrm{d}r \beta^{'2} \beta^{2} \ .
\eeq
For $\lambda_2>\lambda$, this clearly lowers the energy and is thus an unstable direction. 
Depending on the gauge coupling $g$ and the condensate self-couplings $\lambda_1$ and $\lambda_2$, there are regions in parameter space where the global vortex solution is unstable and decays immediately, but we could also identify regions of meta-stability, see Figures \ref{fig:parameter_space_scan_plot} and \ref{fig:parameter_space_contour_plot}. 
In Figure \ref{fig:parameter_space_contour_plot}, the solid line corresponds to the instability boundary in parameter space as derived from the perturbation (\ref{eq:unstable-mode}), which seems to hold approximately for small gauge couplings.

\subfiglabelskip=0pt
The dashed line in Figure \ref{fig:parameter_space_contour_plot} corresponds to the region in parameter space where $\lambda_1=\lambda_2$. This has been identified as the physically relevant regime at ultra-high density, where the coupling is sufficiently small to allow for mean-field calculations, \cite{Iida:2000ha,Giannakis:2001wz},
\beq \label{LambdaExpressions}
\lambda_{1} = \lambda_{2} = \dfrac{\lambda}{4} = \dfrac{36}{7} \dfrac{\pi^4}{\zeta(3)} \left(\dfrac{T_{c}}{\mu} \right) ^{2}.\eeq 
If this result can be extrapolated to densities where the system is strongly coupled, our study indicates that there are no regions of metastability for the physically relevant case. This is supported by the fact that the region of metastability dies away quickly with increasing coupling $g$. 

\section{Initial condition and boundary treatment for multi-vortex dynamics}
\label{sec-1}
\begin{figure}[h]
\centering
\includegraphics[width=7cm,clip]{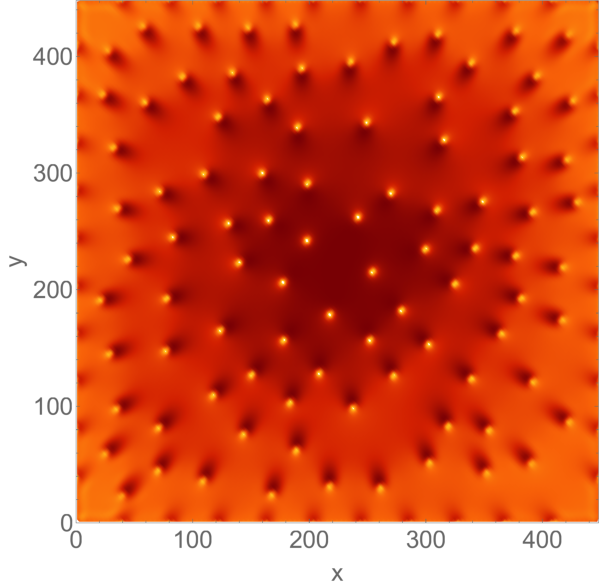}
\caption{Example of the energy density of an initial configuration of 90 vortices. The vortex positions have been chosen randomly, with the additional constraint of a minimal vortex separation of 35 lattice units. In addition to the 90 vortices, one can see the vortices sitting \textit{beyond} the simulation region. They appear as 'shadows` along the edges of the lattice. Those vortices have been frozen into the fixed boundary condition at initial time, and help in providing a better initial setting for large vortex number arrangements, see text.}
\label{fig-multi_vortex_init}
\end{figure}

So far we have explored the behavior of a single global vortex, as well as arrangements of three well-separated semi-superfluid flux tubes. Currently we are investigating two possible extensions to our numerical framework. On the one hand, we introduced strange-quark mass asymmetry, where we followed the work of \cite{Eto:2009tr}. Our findings will be presented in a future publication. On the other hand, we study the dynamics of multiple vortices on the lattice, which introduces several complications. In these proceedings, we discuss the first steps in the simulation of multiple vortices on the lattice.    
To set up many vortices on the lattice, we first obtain a radial profile for a winding-one global vortex by solving (\ref{eq:radial_profile}) using a relaxation method on a large grid with high precision. In the absence of a gauge field at initial time (that is, the link variables are unity), we construct the $\Phi$-field in the presence of $N$ vortices according to
\beq \label{eq:multi_vortex_ansatz}
\Phi = \left(\prod_{i=1}^N \beta(r_i)\right)\exp\left\{i\sum_{i=1}^N n_i\theta_i\right\}\bar\phi\mathbf{1}_{3\times 3},
\eeq
which is just a superposition of the $N$ vortices. At a given point $(r,\theta)$, the  variable $r_i$ indicates the distance to a vortex at position $i$, and $\theta_i$ is the angle with respect to the vortex at position $i$.
Even though we plan to simulate systems of multiple vortices at non-zero gauge coupling in the unstable regime, we start with the most simple case of zero gauge coupling and in the meta-stable region of couplings. In this case, the Ansatz (\ref{eq:multi_vortex_ansatz}) is all we need. Since the superfluid vortices feature a repulsive behavior, they separate quickly once we start our simulation, and, if we use the boundary conditions of our previous study, approach the boundary and pin there. As far as semi-superfluid strings are concerned, the fixed boundary case of our initial work should serve our purpose, since the boundary is repulsive to the individual flux tubes. Together with the repulsion of the semi-superfluid vortices among themselves, this will allow us to find the lattice structure of the multi-fluxtube arrangement. However, at this point we have not yet observed the decay into semi-superfluid fluxtubes in the multi-vortex system, and in this first step we intend to study the numerically much cheaper approach of global vortices only. We thus have to use a modification of the first boundary condition. Using fixed boundary conditions without any modifications also introduces the problem that, far away from the center, many vortices constitute a large overall winding number, which in turn causes the energy density to increase with decreasing distance to the boundary. In order to cancel some of the winding, we introduced a grid of vortices which resides \textit{beyond} the boundary, that is, outside of the simulation region. The presence of those vortices, however, is taken into account when the initial condition is computed, and, in particular, is frozen into the fixed boundary, see Figure \ref{fig-multi_vortex_init}.   
  
\section{Time evolution of a multi-superfluid-vortex arrangement}
\label{sec-2}
\begin{figure*}
\centering
\subfigure[][]{
 \label{fig:multi_vortex_sim_plot_a}
\includegraphics[width=0.45\hsize]{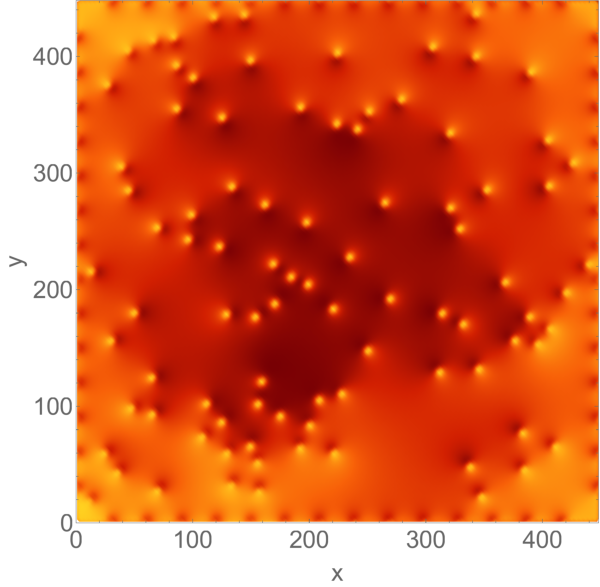}
}\hspace{8pt}
\subfigure[][]{
 \label{fig:multi_vortex_sim_plot_b}
\includegraphics[width=0.45\hsize]{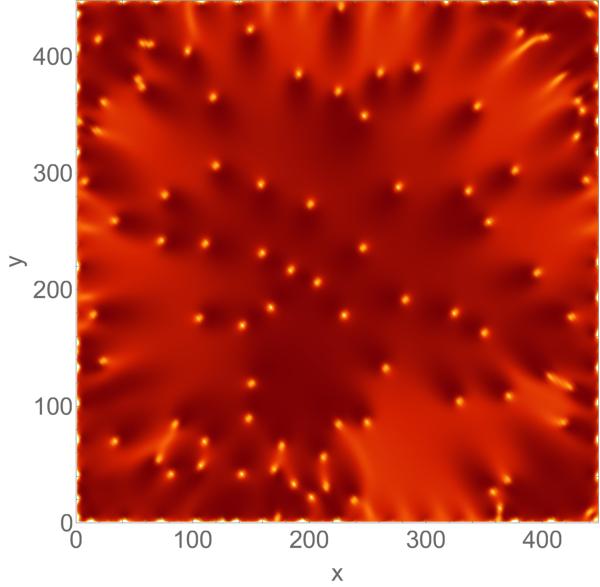}
}\\
\subfigure[][]{
 \label{fig:multi_vortex_sim_plot_c}
\includegraphics[width=0.45\hsize]{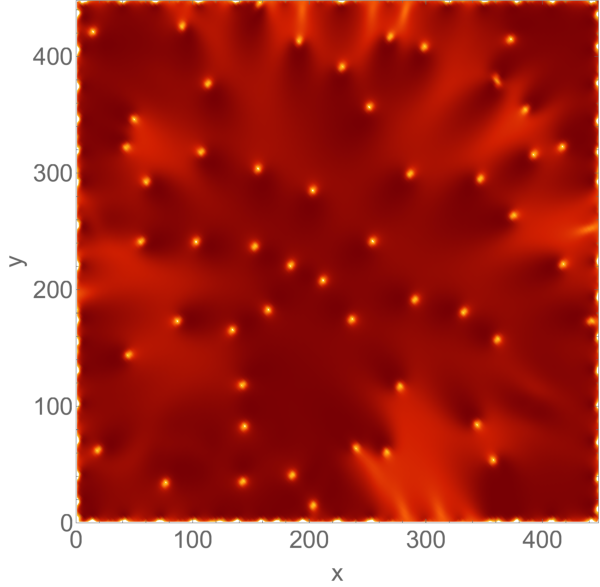}
}
\hspace{8pt}
\subfigure[][]{
 \label{fig:multi_vortex_sim_plot_d}
\includegraphics[width=0.45\hsize]{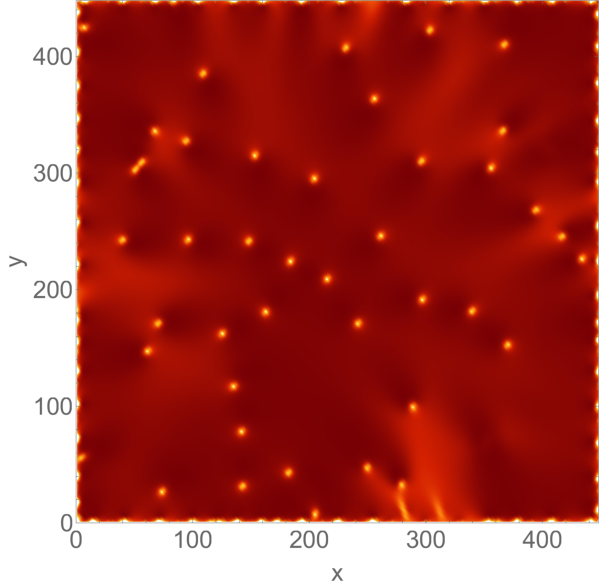}
}
\caption[]{Energy density snapshots of the time evolution of an initial arrangement of 90 vortices at random positions with minimal initial vortex-vortex separation of 15 lattice units, using the improved fixed boundary condition discussed in Section \ref{sec-2}. The panels \ref{fig:multi_vortex_sim_plot_a},\ref{fig:multi_vortex_sim_plot_b},\ref{fig:multi_vortex_sim_plot_c} and \ref{fig:multi_vortex_sim_plot_d} show the state of the system at initial time, and at $t=1000$, $t=2000$ and $t=3000$ respectively. In order to prevent a violent separation of the vortices, a high damping factor has been used in this evaluation. The lighter color shading of the initial state in panel \ref{fig:multi_vortex_sim_plot_a} comes from the fact that the initial condition has a lot of excess energy, which is then dissipated throughout the time-evolution.}
\label{fig:multi_vortex_sim_plot}
\end{figure*}

In this section we show the time evolution of a random setting of 90 vortices on a $448\times448$ lattice at vanishing gauge coupling in the meta-stable regime of the theory. In this case, the superfluid vortices won't undergo the decay to the semi-superfluid fluxtubes, but are expected to form a hexagonal lattice structure. Snapshots of the time evolution of a first simulation run are shown in Figure \ref{fig:multi_vortex_sim_plot}.  
\subfiglabelskip=0pt
In this run we have used a very high damping factor, in order to relax the system towards its lowest energy state as gentle as possible. This is necessary, since the initial state has a high vortex (winding) number, and since our Ansatz is not the lowest energy state, it has a lot of excess energy. The downside of a high damping factor is a very slow time evolution, which is evident from the comparably small changes in the configurations shown in Figure \ref{fig:multi_vortex_sim_plot_a}, despite the large time separation of the configurations. In this first simulation run we successfully managed to simulate the time evolution of a state with very high vorticity in a controlled and stable way. The vortices separate, and seem to prefer to occupy the boundary, which requires some improvement of the boundary condition. So far we did not observe the expected hexagonal lattice structure, however, this should be easily achievable by an appropriate choice of initial condition and simulation parameters.
\section{Conclusions and outlook}
\label{sec-3}
In this progress report we have reviewed our findings from our previous study \cite{Alford:1997zt}, where we have found numerical proof of the instability of superfluid vortices in CFL quark matter. We identified regions of stability and metastability in the parameter space of the couplings, and connected our result to the physically relevant regime by extrapolating weak coupling results. This indicates that superfluid vortices are unstable in CFL quark matter. An analysis of the coupling dependence of the stability/metastability boundary revealed that it seems to be independent of the coupling $\lambda_2$, which raises the question of the role of the condensate self-couplings in the decay process. We also constructed a mode that is sufficient to trigger the decay of a superfluid vortex into the triplet of semi-superfluid fluxtubes. The phenomenological consequences of the instability, for example on neutron star glitches, remain elusive for now, since that requires a thorough understanding of a pinning mechanism. As a first step towards a better understanding of the dynamics of a system with high vorticity we extended our framework and presented first results for the time evolution of a multi-superfluid-vortex system. We modified the boundary conditions slightly, which allows for a better initial setting of a multi-vortex state, and simulated a system of 90 superfluid vortices in a controlled and stable way. Because it is numerically less expensive, we started with the zero-gauge coupling case, and focused on the stable regime of condensate self-couplings. As a next step, we intend to study the non-zero gauge coupling case. This is numerically more expensive, since, on the lattice, gauge fields are represented as link variables corresponding to group elements, and their time evolution involves a matrix exponential. In order to speed up the code, we thus plan to use Graphics Processing Units (GPUs), which allows for much faster evolution on larger lattices.

\section*{Acknowledgments}
We thank Andreas Schmitt for helpful discussions. This study has been supported by the U.S. Department of Energy, Office of Science, Office of Nuclear Physics under Award number \#DE-FG02-05ER41375. AW acknowledges support through the Austrian Science Fund (FWF) Schr\"odinger Fellowship J 3800-N27. TV acknowledges support by the U.S. Department of Energy, Office of Science, Office of High Energy Physics under Award number \#DE-SC0013605.

%
\bibliography{vortices}

\end{document}